\documentclass[%
reprint,
superscriptaddress,
amsmath,
amssymb,
pra,
]{revtex4-1}

\usepackage{graphicx} 

\usepackage{mathptmx} 
\usepackage{xcolor}   
\usepackage{physics}  
\usepackage{dcolumn}
\usepackage{tabularx}
\usepackage{tikz}
\usetikzlibrary{arrows}
\usepackage{caption}
\usepackage{subcaption}
\usepackage{lipsum}
\usepackage{overpic}
\usepackage{svg}
\usepackage[export]{adjustbox}

\usepackage{hyperref} 
\hypersetup{
    colorlinks=true,
    linkcolor=blue,
    citecolor=blue,
    filecolor=blue,
    urlcolor=blue
}

\begin{document}


\author{Rajavardhan Talashila}
\altaffiliation[These authors contributed equally to this work]{}
 \affiliation{Department of Electrical Engineering, University of Colorado, Boulder, Colorado 80302, USA}
\affiliation{Associate of the National Institute of Standards and Technology, Boulder, Colorado 80305, USA}
\author{William J. Watterson}
 \altaffiliation[These authors contributed equally to this work]{}
 \affiliation{Department of Physics, University of Colorado, Boulder, Colorado 80302, USA}
\affiliation{Associate of the National Institute of Standards and Technology, Boulder, Colorado 80305, USA}
\author{Benjamin L. Moser}
\author{Joshua A. Gordon}
\author{Alexandra B. Artusio-Glimpse}
\author{Nikunjkumar Prajapati}
\author{Noah~Schlossberger}
\author{Matthew T. Simons}
\author{Christopher L. Holloway}
\email{christopher.holloway@nist.gov}
\affiliation{ National Institute of Standards and Technology, Boulder, Colorado 80305, USA}

\date{\today}

\title{Determining angle of arrival of radio frequency fields using subwavelength, amplitude-only measurements of standing waves in a Rydberg atom sensor}

\begin{abstract}

Deep subwavelength RF imaging with atomic Rydberg sensors has overcome fundamental limitations of traditional antennas and enabled ultra-wideband detection of omni-directional time varying fields all in a compact form factor. However, in most applications, Rydberg sensors require the use of a secondary strong RF reference field to serve as a phase reference. Here, we demonstrate a new type of Rydberg sensor for angle-of-arrival (AoA) sensing which utilizes subwavelength imaging of standing wave fields. By placing a metallic plate within the Rydberg cell, we can determine the AoA independent of the strength of incoming RF field and without requiring a secondary strong RF phase reference field. We perform precision AoA measurements with a robotic antenna positioning system for 4.2, 5.0, and 5.7 GHz signals and demonstrate a $1.7^{\circ}$ polar angular resolution from $0^{\circ}$ to $60^{\circ}$ AoA and $4.1^{\circ}$ over all possible angles.

\end{abstract}

\maketitle

\section{Introduction}

Determining the angle of arrival (AoA) of an incoming radio-frequency (RF) signal is typically achieved by receiving the signal at multiple locations and using knowledge of the amplitude and/or phase differences to determine the AoA through various algorithms/methods. The resolution achieved by these methods depends on the number of elements used and the stability of the phase reference. This places limits on the size of a receiving device and how closely individual antennas can be spaced. As opposed to traditional antennas, Rydberg atom-based electric field measurements can be made at a spacing that is fractions of a wavelength without interfering with adjacent measurements \cite{Noah2024}.
Rydberg atom field measurements are based on an incident RF field perturbing the atomic state of a Rydberg atom~\cite{gallagher_book}. By detecting the change of the energy states of the atom, the magnitude and polarization of the incident RF field~\cite{Noah2024,9748947, simons_rydberg_2021} can be determined, and introducing a local oscillator (LO) the phase can also be determined. Previous work demonstrated that a Rydberg atom sensor can be used to determine AoA of an incident field~\cite{AOA, 10304615, yan2023three}, where the phase of an incident RF field is resolved at different locations inside a vapor cell (a glass cell that contains the atomic vapor). The phase difference between multiple locations is used to determine the AoA of the field. The phase of the incident field is determined through the use of a Rydberg atom mixer~\cite{doi:10.1063/1.5088821}. In this approach, an additional RF field is applied to the vapor cell and acts as a local oscillator. The two fields beat against each other and the phase of the beat note is related to the relative phase of the incident field and the LO.

The limitations of this approach are that  it requires an additional RF field (the LO) and reflections from the local environment can destroy the connection between phase difference and AoA. Here we present an approach that does not require an additional LO field, and is based on the standing wave pattern that occurs inside a vapor cell with  an integrated metal plate, which can be approximated as a perfect electric conductor (PEC). This simple device we term a PEC-Rydberg cell. The vapor cell is made of glass (a dielectric) and as such, weak standing waves (an interference pattern) can develop inside the vapor cell~\cite{holl3,PhysRevApplied.4.044015}. The addition of the metal plate allows for a tailored reflection pattern and reduces the effect of unwanted reflections from the local environment. The standing wave pattern inside the vapor cell is a function of the angle of the incident RF signal. This allows us to determine the AoA via amplitude-only measurements, such that no phase information is required. 



\begin{figure*}[!ht]
    \includegraphics[width=0.95\linewidth]{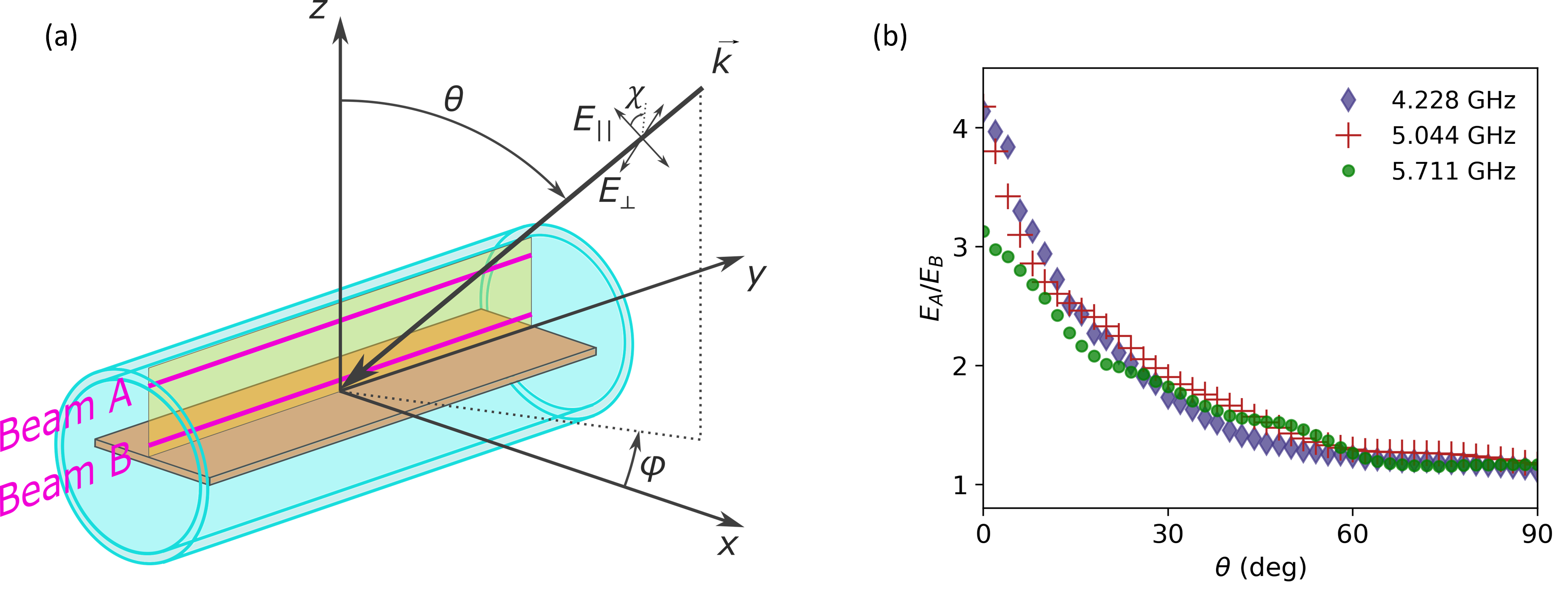}
   
    \caption{\textbf{The PEC-Rydberg antenna uniquely determines incoming angle-of-arrival (AoA) of an incoming RF field through amplitude-only, subwavelength RF imaging of standing wave field patterns.} (a) Illustration of the PEC-Rydberg antenna. The PEC plate (orange) runs along a cylindrical vapor cell's major axis in the xz-plane. Two E-fields are measured through Rydberg Autler-Townes splitting along the two pink lines in the yz-plane. An E-field is incident with coordinates $(\theta, \phi, \chi)$, where $\theta$ is the rotation from the z-axis, $\phi$ is the rotation from the x-axis, and $\chi$ is the polarization defined as the angle from the plane of incidence. (b) The measured ratio of the electric field along Beams A and B, $E_A/E_B$, decreases monotonically with increasing $\theta$ from 4.2 to 5.7 GHz, thereby providing a unique determination of AoA at each frequency point.}
    \label{Fig:Intro}
\end{figure*}

\begin{figure*}[!ht]
    \includegraphics[width=\linewidth]{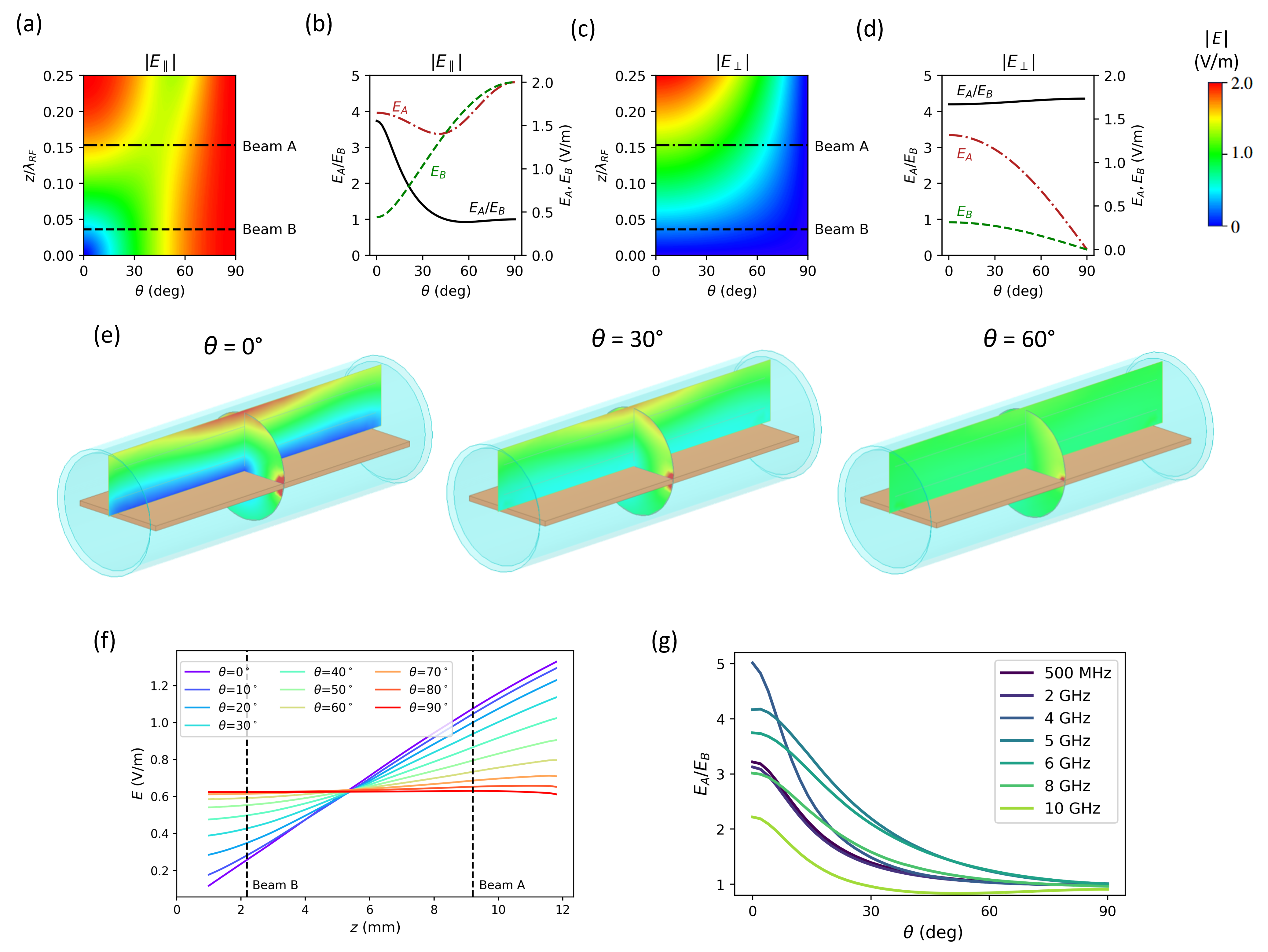}
    
    \caption{\textbf{Analytical and numerical simulations indicate the ratio of two field strengths, $E_A/E_B$, measured inside the PEC-Rydberg cell can determine the incoming AoA.} (a-d) The standing wave field pattern at a normalized distance, $z/\lambda_{RF}$, as a function of incoming angle-of-arrival (AoA), $\theta$, for an incident 1.0 V/m E-field incident on an infinite conducting plane polarized (a) parallel and (c) perpendicular to the plane of incidence. The laser beam locations we test for RF sensing are indicated by the dashed lines. E-fields along Beam A and B, $E_A$ and $E_B$, respectively, and the ratio of the fields, $E_A/E_B$, as a function of $\theta$, for fields polarized (b) parallel and (d) perpendicular to the plane of incidence. (e) Finite element simulation results showing the $E_{\parallel}$ field distribution in the the vapor cell in the yz-plane and xz-plane for three different angles of incidence. (f) Numerically determined average field strength along the length of the cell for varying $\theta$. Beam A and B location are denoted by the black dashed lines. (g) The field ratio, $E_A/E_B$, for varying frequency and $\theta$.}
    \label{Fig:Simulation}
\end{figure*}

\begin{figure}[!ht]
    \captionsetup[subfigure]{labelformat=empty}
    \begin{flushleft}
    \subfloat[\label{Fig:Transitions}]{
      \resizebox{5cm}{!}{
        \begin{tikzpicture}[
          scale=0.5,
          level/.style={thick},
          virtual/.style={thick},
          trans/.style={thick,->,shorten >=2pt,shorten <=2pt,>=stealth}
        ]\flushleft
        \draw[level] (2cm,0em) -- (0cm,0em) node[left] {6S\textsubscript{1/2} (F=4)};
        \draw[level] (2cm,6em) -- (0cm,6em) node[left] {6P\textsubscript{3/2} (F=5)};
        \draw[level] (2cm,12em) -- (0cm,12em) node[left] {nD\textsubscript{5/2}};
        \draw[level] (4cm,8em) -- (2cm,8em) node[below right] {(n+1)P\textsubscript{3/2}};
        \draw[trans] (1cm,0em) -- (1cm,6em) node[midway,left, red] {850 nm};
        \draw[trans] (1cm,6em) -- (1cm,12em) node[midway,left, green!90!black!90] {512 nm};
        \draw[trans] (1cm,12em) -- (3cm,8em) node[midway, right] {RF};

        \node[] at (-3,4) {(a)};
        \end{tikzpicture}
        }
    }
    \end{flushleft}

    \vspace{-1.8cm}
    
    \subfloat[\label{Fig:ExperimentSetup_b}]{%
      \begin{overpic}[width=0.8\linewidth]{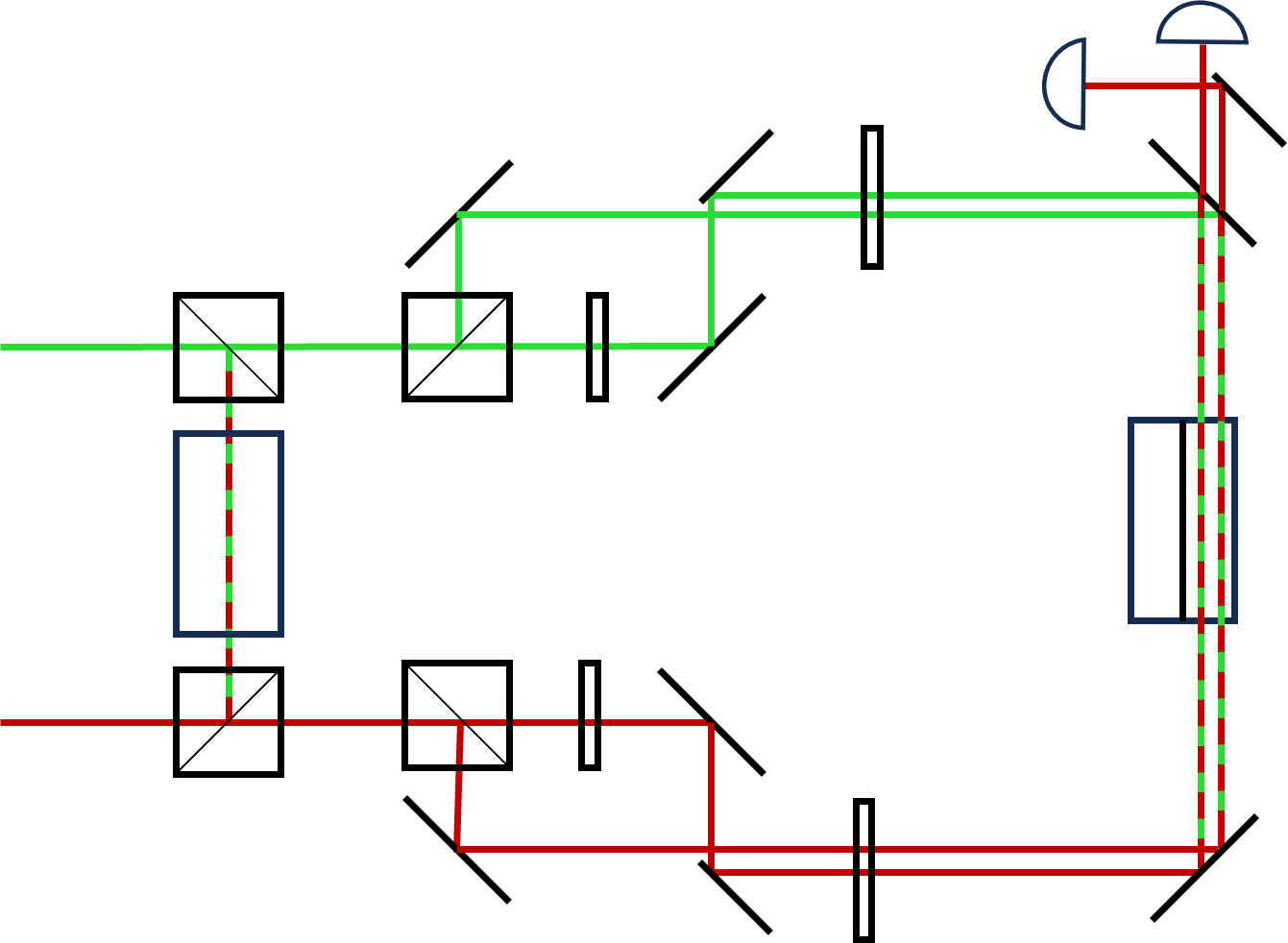}
        \put(0,18){850 nm}
        \put(0,47){512 nm}
        \put(43,23){$\lambda$/2}
        \put(43,38){$\lambda$/2}
        \put(64,12){$\lambda$/4}
        \put(64,49){$\lambda$/4}
        \put(96,57){DM}
        \put(77,70){PD}
        \put(93,74){PD}
        \put(5,24){\rotatebox{90}{Reference}}
        \put(9,28){\rotatebox{90}{Cell}}
        \put(83,25){\rotatebox{90}{Test Cell}}
        \put(2,3){(b)}
      \end{overpic}
    }
    
  \subfloat[\label{Fig:ExperimentSetup_a}]{%
      \begin{overpic}[width=0.85\linewidth]{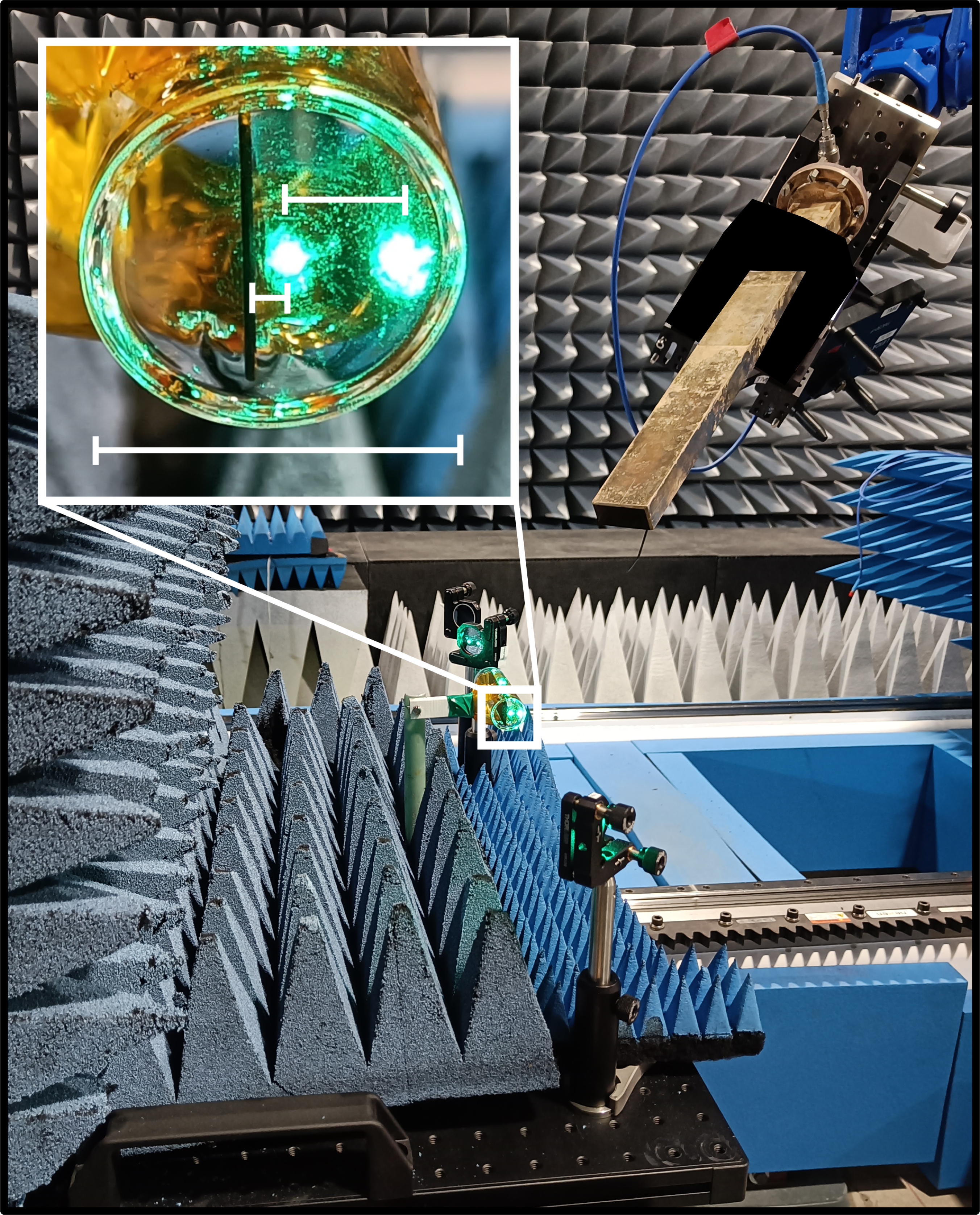}
        \put (2,3){\colorbox{white}{\makebox(3,3)[c]{(c)}} }
        \put(20,60){\color{white} 25 mm}
        \put(21,70){\color{white} 2.2 mm}
        \put(24,86){\color{white} 7 mm}
        
      \end{overpic}
    }
    
    \caption{\textbf{Experimental setup for measuring the AoA of an incoming CW wave.} (a) Cesium level diagram for RF detection. (b) Optical schematic: half-wave plate ($\lambda/2$), quarter-wave plate ($\lambda/4$), photodetector (PD), dichroic mirror (DM). Mirrors are indicated by diagonal lines and polarizing beam splitters by squares with an internal diagonal. (c) The vapor cell is shown on the portable optical setup inside the LAPS facility at NIST. The WR187 open-ended waveguide is at coordinates $(\theta, \phi, \chi)$ = (45$^{\circ}$, -45$^{\circ}$, 0$^{\circ}$) and at a distance of 20 cm. The inset shows a zoomed-in image of the vapor cell with the vertically aligned PEC plate (black line) and the two coupling beams (green spots).}
    \label{Fig:ExperimentalSetup}

\end{figure}

\section{Results\label{sec:Results}}

\subsection{Angle of Arrival Detection}

We begin by highlighting the key data illustrating the PEC-Rydberg sensor AoA detection (Fig.~\ref{Fig:Intro}). We engineered a vapor cell with an integrated metal plate to create a standing wave field pattern that changes as a function of the incoming RF polar and azimuthal coordinates and RF polarization relative to the plane of incidence, $(\theta, \phi, \chi)$, respectively (Fig. \ref{Fig:Intro}a). The amplitude of the electric field is measured along two paths parallel to the PEC plate through resonant Autler-Townes splitting in Rydberg atoms of Cs\textsuperscript{133} (See Methods). This Rydberg-based RF imaging allows a non-perturbative and subwavelength probing of the standing wave field structure, which can be directly related to the incoming AoA as described later. In particular, we determine that the ratio of E-field amplitude along the beam at two locations, $E_A/E_B$, gives a monotonic determination of the incoming AoA. A representative dataset from 4.2 to 5.7~GHz for $(\phi, \chi)=(0^{\circ}, 0^{\circ})$ shows the one-to-one mapping of $E_A/E_B$ versus AoA, $\theta$ (Fig.~\ref{Fig:Intro}b). 

Importantly, we note this method requires sub-wavelength sampling to determine the absolute field strength and is therefore not possible for traditional antennas because of size limitations and perturbation of the field by each antenna. Rydberg atom-based field measurements, on the other hand, do not interfere, allowing multiple measurements to be made on a sub-wavelength scale~\cite{fan_effect_2015,holloway_sub-wavelength_2014}. 

The basic principle behind the operation of our PEC-Rydberg cell can be understood by examining the standing wave patterns of an electric field incident on an infinite conducting plane (Fig.~\ref{Fig:Simulation}a-d). A plane wave of wavenumber magnitude, $k_{RF} = 2\pi / \lambda_{RF}$, incident on an infinite perfectly conducting plane with angle of incidence, $\theta$, can be analyzed in the two cases of the vector electric field being parallel~($\parallel$) ($\chi = 0)$ and perpendicular~($\perp$) ($\chi = \pi/2$) to the plane of incidence \cite{popović2000introductory,zahn2003electromagnetic}. The amplitudes of the total electric field in both cases can be expressed as

\begin{equation}
\centering
\begin{split}
     |E_{\parallel}| &\propto \sqrt{ 1 - \cos(2\theta)\cos(2k_{RF}z\cos(\theta)) }\\
     |E_{\perp}| &\propto \sqrt{1 -\cos(k_{RF}z\cos(\theta)) },
     \label{AnalyticEquation}
\end{split}
\end{equation}
where $E_{\parallel}$ is the total electric field amplitude for the parallel case and $E_{\perp }$ is the total electric field amplitude for the perpendicular case. At normal incidence, $E_{\parallel}$ has a node ($E_{\parallel}=0$ V/m) at the surface of the conductor ($z=0$) and an anti-node at a distance $\lambda_{RF}/4$ away (Fig.~\ref{Fig:Simulation}a).  As the angle of incidence is increased, the standing wave pattern changes its characteristics with respect to the distance $z$ from the conducting plate. It can be noticed that the ratio of the field strength at two locations near the plate, $E_{\parallel}(z_A)/E_{\parallel}(z_B)$, gives a near-monotonic behaviour (Fig.~\ref{Fig:Simulation}c) enabling the determination of the incoming angle independent of absolute field strength of the plane wave incident. In the case of $E_{\perp}$, while $E_{\perp}(z_A)/E_{\perp}(z_B)$ is also monotonic over $\theta$, the ratio does not vary significantly so any uncertainty in the measured fields will lead to a large uncertainty in the incoming angle.

The closed-form solution of the infinite conducting plane (Eq.~\ref{AnalyticEquation}) affords an intuitive discussion of the basic AoA principle used; however, in our PEC Rydberg cell, the finite conducting plane enclosed in a dielectric vapor cell must be modeled numerically (see Experimental Methods). The relevant geometry and coordinate system is illustrated in Fig.~\ref{Fig:Intro}a. The standing wave field strength, $\abs{E_{\parallel}}$, as a function of incoming angle for 5.044~GHz and $\chi = 0^{\circ}$ is shown in Fig.~\ref{Fig:Simulation}e-f. We observe at $\theta = 0^{\circ}$, $\abs{E}$ = 0~V/m at the PEC plate in the yz-plane and increases to an anti-node along the z-axis. However, the dielectric edges of the cylindrical vapor cell cause field distortions relative to the infinite conducting plane case. As $\theta \rightarrow 90^{\circ}$, the field strength $E_{\parallel}(x,z_B) \rightarrow 1$. The ratio of the field strengths along the length of the cell at two locations (A and B) demonstrates the monotonic determination of AoA using the PEC-Rydberg cell, which  extends from 500~MHz to 9~GHz (Fig.~\ref{Fig:Simulation}g) for the dimensions of the cell considered in this work. 

We experimentally test the AoA principle using the PEC-Rydberg cell by measuring the field amplitude at two locations, A and B, through resonant Autler-Townes (AT) splitting in Rydberg states of Cs\textsuperscript{133} \cite{holloway_broadband_2014} (Methods, Fig.~\ref{Fig:ExperimentalSetup}a). We designed a portable optical setup and performed angle-of-arrival measurements for varying $(\theta, \phi, \chi)$ from 4.2~GHz to 5.7~GHz at the Large Antenna Positioning System at NIST in Boulder, Colorado (Methods, Fig.~\ref{Fig:ExperimentalSetup}b-c).

\begin{figure*}[t]
    \includegraphics[width=0.9\linewidth]{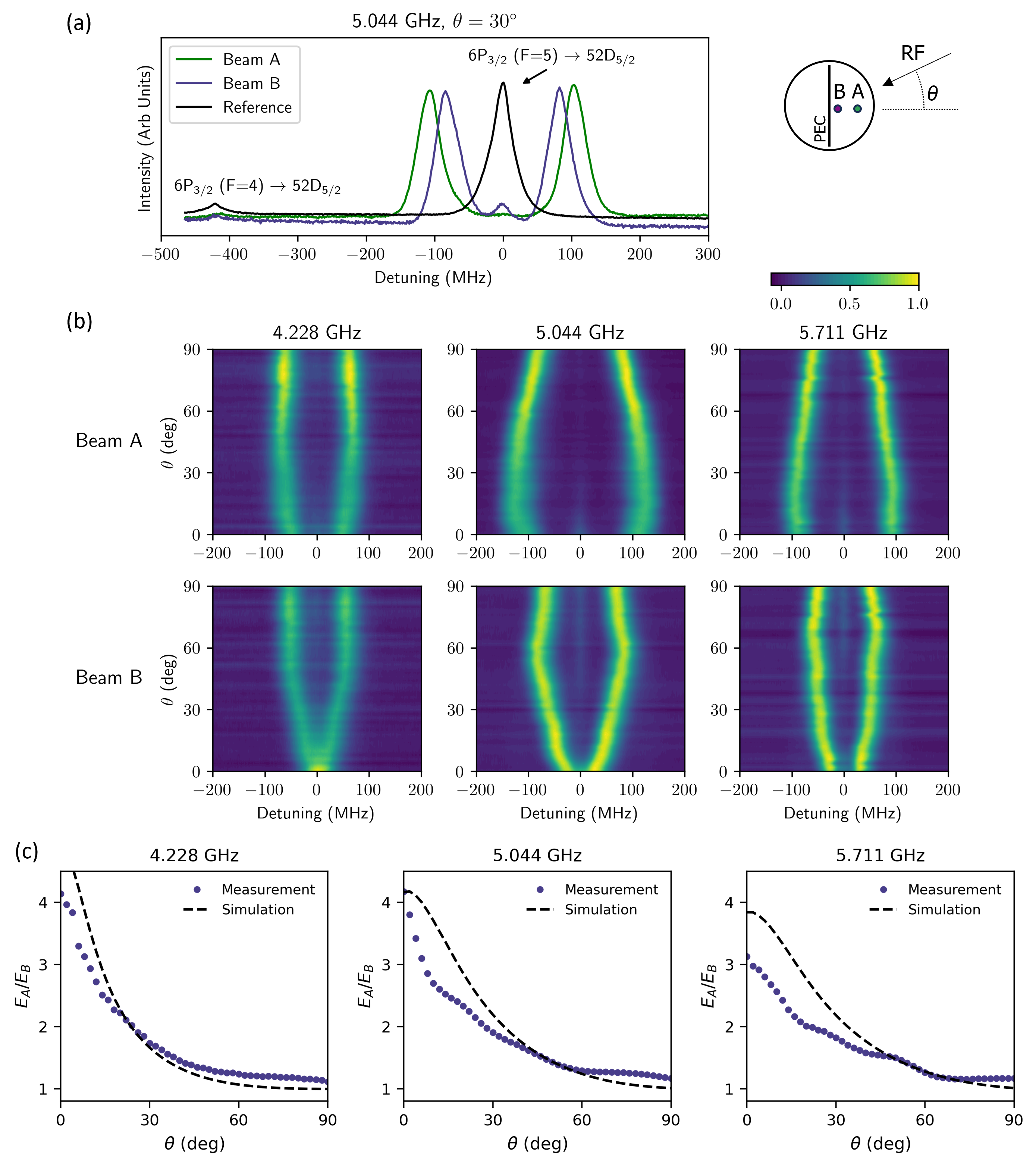}
    \caption{\textbf{The AoA is uniquely determined by our PEC-Rydberg vapor cell from 4.2 to 5.7~GHz.} (a) EIT traces in the PEC-Rydberg cell and reference cell. The cartoon to the right indicates the beam locations, A and B, with respect to the PEC plate and the incoming RF field. For Beam A and B, the EIT with Autler-Townes splitting is shown for a 5.044 GHz field with $\theta = 30 ^{\circ}$. (b) Normalized EIT signals in Beam A and B (rows) for $f_{RF}$ = 4.228~GHz, 5.044~GHz, and 5.711~GHz (columns) for varying $\theta$. (c) A one-to-one mapping exists between $\theta$ and $E_A/E_B$ for each tested frequency indicating the PEC-Rydberg cell can be calibrated to uniquely determining the incoming AoA.}
    \label{Fig:AoA}
\end{figure*}

\begin{figure}[t]
    \includegraphics[width=0.9\linewidth]{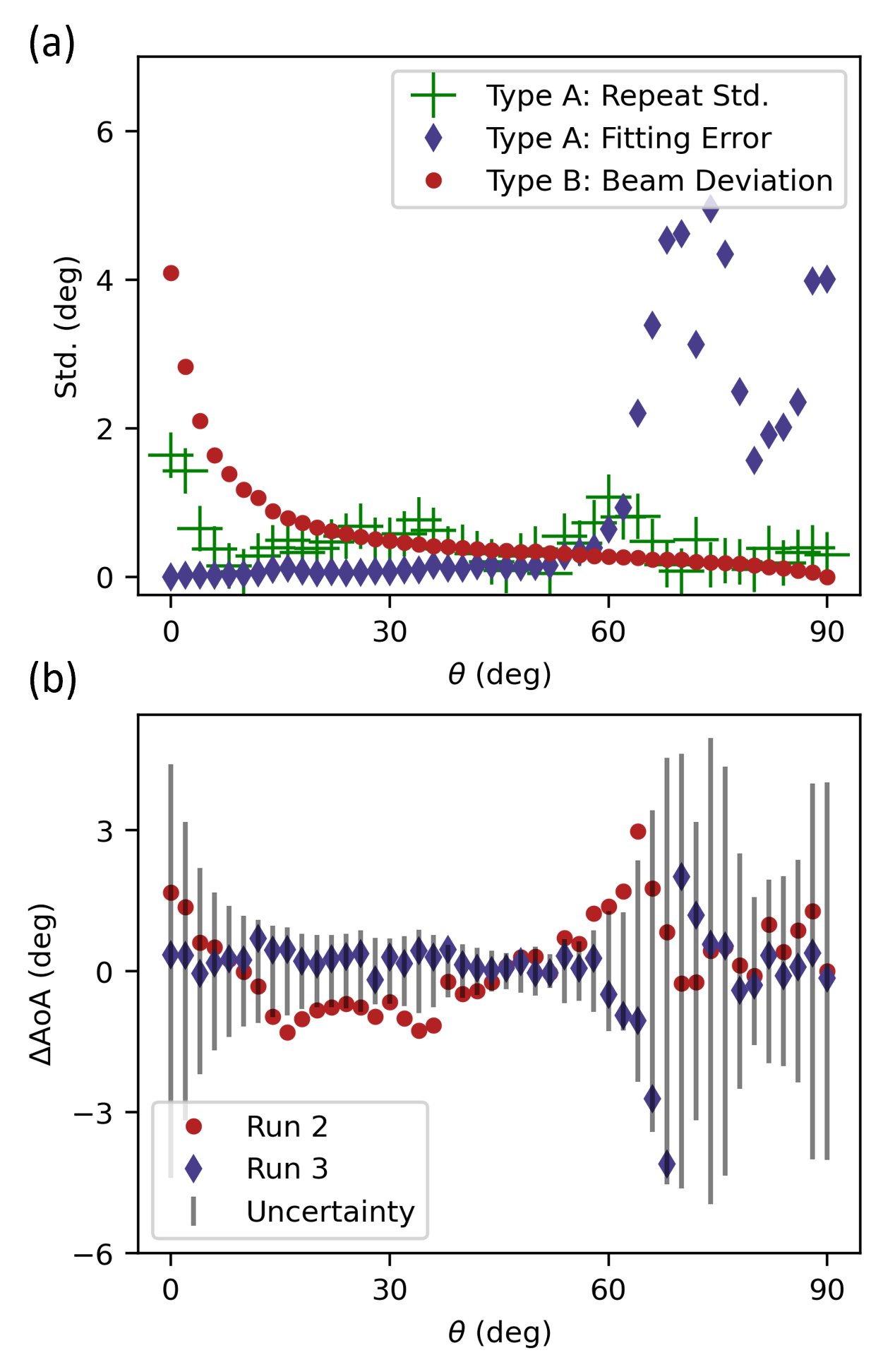}
    \caption{\textbf{The maximum deviation from 5 GHz measurement repeats is 1.7$^{\circ}$ for $\theta \leq 60^{\circ}$ and less than 4.1$^{\circ}$ across all $\theta$.} (a) Components of uncertainty as a function of incoming angle. (b) Three repeat measurements of 5.044 GHz were obtained and one measurement is used as a calibration to determine the AoA. The uncertainty bars are obtained from the square root of the sum of squares of the uncertainties in (a).}
    \label{Fig:Resolution}
\end{figure}



        

\begin{figure*}
    \includegraphics[width=\textwidth]{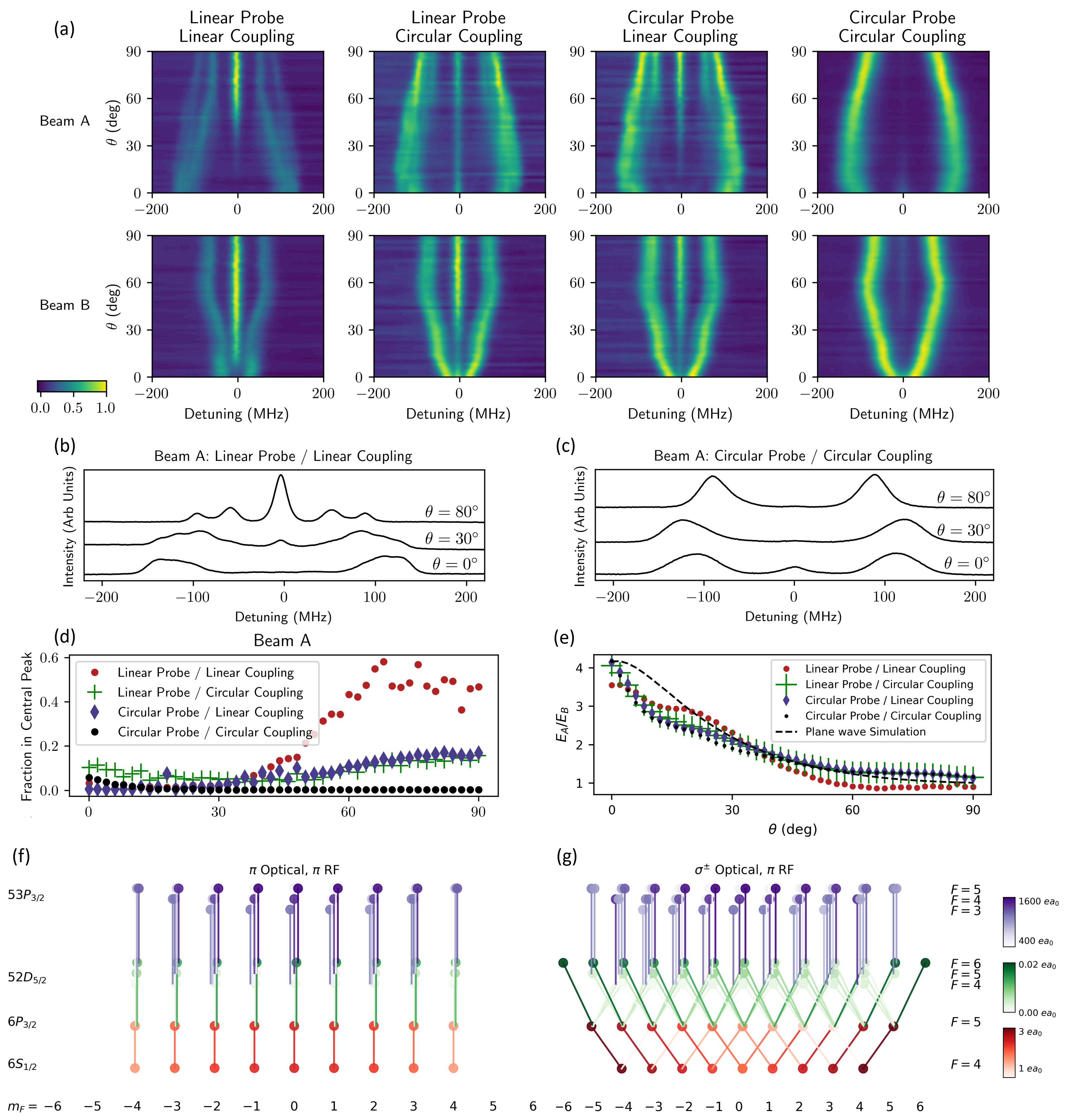}%
    \caption{\textbf{Circularly polarized probe and coupling beams best resolve the incoming AoA.} (a) Normalized EIT signals in Beam A and Beam B for different combinations of probe and coupling optical field polarization. One dimensional traces in Beam A of the EIT signal for (b) linearly polarized probe and coupling and (c) circularly polarized probe and coupling. (d) The fraction of the population in the central peak for both Beam A (e) The ratio, $E_A/E_B$, for the different optical field polarizations. Excitation transition state diagrams for (f) $\pi$ polarized probe, coupling, and RF fields (left) and (g) $\sigma^{\pm}$ polarized probe and coupling with $\pi$ polarized RF. The transition dipole moments for each transition are given by the colormaps to the right, where $e$ is the electron charge and $a_0$ is the Bohr radius. 
    }
    \label{Fig:Polarization}
\end{figure*}

\begin{figure}[htbp]
\centering
\includegraphics[width=0.9\linewidth]{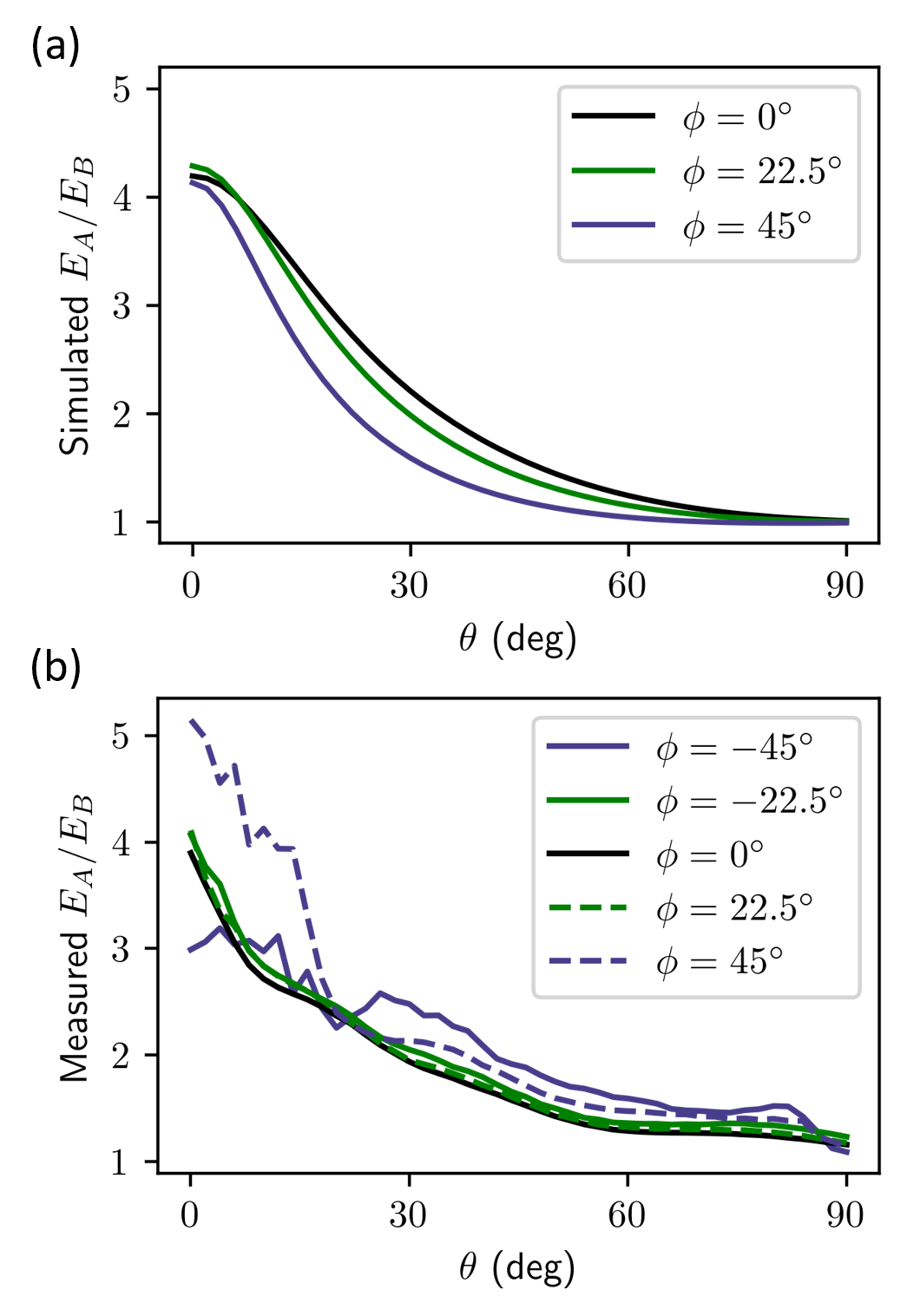}
\caption{\textbf{The AoA measurement can be well resolved within a 45$^{\circ}$ range of varying azimuthal angle.} The ratio, $E_A/E_B$, for varying $\phi$ (a) simulated and (b) measured. At $\phi = \pm 45^{\circ}$, reflections in the nearby optical mounts caused significant noise in the measured ratio.}
\label{Fig:Phi}
\end{figure}

\begin{figure*}[htbp]
\centering
\begin{overpic}[width=\linewidth]{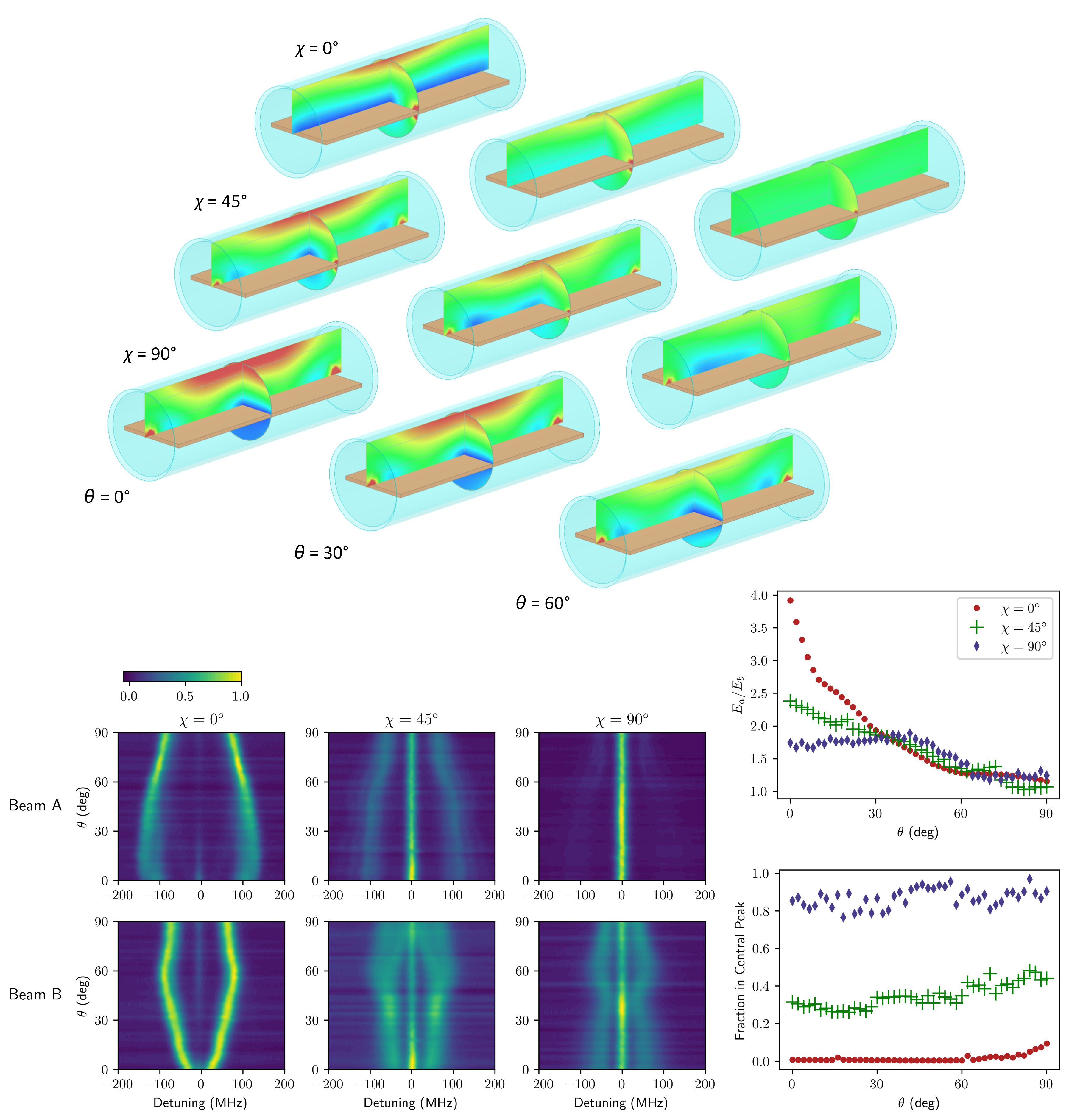}
        \put(2,95){(a)}
        \put(2,38){(b)}
        \put(64,47){(c)}
        \put(64,23){(d)}
        \put(83,53){\includegraphics[width=0.018\linewidth]{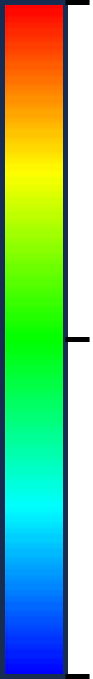}}
        \put(85,52.5){\footnotesize 0}
        \put(85,59){\footnotesize 1.0}
        \put(85,65){\footnotesize 2.0}
        \put(82.5,69){$\abs{E}$}
        \put(81.5,67){(V/m)}
\end{overpic}

\caption{\textbf{The fraction of the EIT signal in the central peak in Beam A determines the RF polarization and enables AOA measurements from $\chi = 0^{\circ}$ to $45^{\circ}$.} (a) Finite element simulations of the field magnitude, $\abs{E}$, in the xz-plane and yz-plane for varying $\theta$ and $\chi$. The upper gray line in each vapor cell indicates the Beam A path and the lower line the Beam B path. (b) Normalized EIT signals in Beam A and Beam B for varying $\theta$ and $\chi$ with both optical beams circularly polarized. (c) The ratio, $E_a/E_b$, for varying $\chi$ demonstrates a one-to-one relationship for $\chi = 0^{\circ}$ and $45^{\circ}$. (d) The fraction in the central peak in Beam A varies from 0 to 1 as $\chi$ varies from $0^{\circ}$ to $90^{\circ}$.}
\label{Fig:RFpolarization}
\end{figure*}

\begin{figure*}[htbp]
\centering
\includegraphics[width=\linewidth]{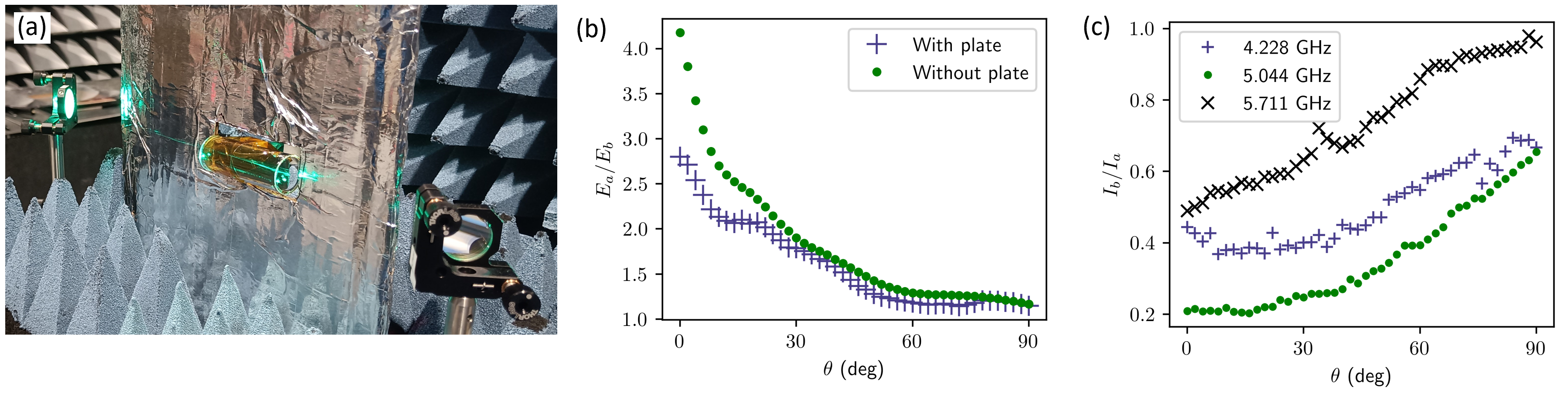}
\caption{\textbf{Paths forward exist towards translating the PEC-Rydberg antenna from laboratory tests to field applications}. (a) The experimental setup of a conducting plane surrounding the vapor cell and in plane with the internal PEC. (b) $E_A/E_B$ under the test setup shown in (a) versus without the plate. (c) Amplitude regime AoA measurement at low RF field strength. The ratio of the intensities in Beam A and B was measured at zero detuning.}
\label{Fig:Applications}
\end{figure*}

We first test the PEC-Rydberg cell varying $\theta=0^{\circ}$ to  $\theta=90^{\circ}$ with $(\phi, \chi)$ = ($0^{\circ}$, $0^{\circ}$). The key features of the collected EIT traces are shown in Fig. \ref{Fig:AoA}a for an example measurement at $\theta = 30^{\circ}$ and $f_{RF}$ = 5.044 GHz. The reference cell shows the main EIT peak from the 6P\textsubscript{3/2} (F=5) $\rightarrow$ 52D\textsubscript{5/2} transition and the hyperfine structure from the 6P\textsubscript{3/2} (F=4) $\rightarrow$ 52D\textsubscript{5/2}. AT splitting is observed for both the beam near the PEC plate, Beam B, and the beam further from the PEC plate, Beam A. The small peak near 0 MHz detuning of the coupling laser frequency in Beam B indicates a subfraction of the Rydberg population that is not undergoing AT splitting due to the relative polarizations of the probe, coupling, and RF fields which we discuss in Fig. \ref{Fig:RFpolarization}. The peaks in Beam A and B are fit to Lorentzian curves to extract the AT peak separations. Fig. \ref{Fig:AoA}b shows the EIT traces represented as heat maps collected at a 2$^{\circ}$ step size for $f_{RF} = $ 4.228~GHz, 5.044~GHz, and 5.711~GHz. The peak separations in Beam A and B were used to obtain the splitting ratio, $E_A/E_B$ for each frequency and compared to the numerical simulations (Fig. \ref{Fig:AoA}c).

The data shown in Fig.~\ref{Fig:AoA}c demonstrates the AoA of the incoming RF fields is uniquely determined by our PEC-Rydberg cell for $(\phi, \chi)$ = ($0^{\circ}$, $0^{\circ}$)  and from 4.2 to 5.7~GHz. Although the EIT splitting in a single channel is not monotonic with varying $\theta$ (e.g. Beam B 5.044~GHz), the ratio of the splittings, $E_A/E_B$ maintains a one-to-one mapping. While the measured data qualitatively matches the numerical simulations, they differ due to fact that the simulation data is obtained by taking into account the E-field magnitude at a single point at the center of the cell along the z-axis. However, in measurement, the lasers sense a non-uniform field along the length of the cell (y-axis) as shown in the field distributions in Fig.~{\ref{Fig:Simulation}}. Such non-uniformity distorts the spectral line shapes \cite{rotunno_investigating_2023} and hence the deviations of the field ratios. Other factors include non-idealities in the exact placement of the PEC plate in the Rydberg vapor cell and possible reflections from optical components and non-ideal absorbers. Thus, the PEC-Rydberg cell would require a calibration before field testing. 

\subsection{Uncertainty Analysis}

In order to estimate the uncertainties of our method, we perform three AoA measurement repeats for 5.044 GHz  across multiple days and experimental runs. Our method consists of three main sources of uncertainty -- Type~A repeatability, Type~A fitting error of the peak locations used to determine $E_A/E_B$, and Type~B deviations in the beam locations relative to each other and PEC plate. We discuss our Type~B beam deviation and other uncertainties further in the Discussion section. As can be seen in Fig.~\ref{Fig:Resolution}a, the beam deviation uncertainties dominate at low $\theta$ and the fitting error uncertainties dominate at high $\theta$. To compare the measurement repeats, we use one measurement as a calibration and  determine the AoA deviation, $\Delta$AoA, for two repeats (Fig.~\ref{Fig:Resolution}b). The maximum deviation between the repeats occurrs at $\theta \simeq 70^{\circ}$ with an absolute angular difference of $4.1^{\circ}$. For lower incidence angles with $\theta \leq 60^{\circ}$, the maximum observed deviation is $1.7^{\circ}$. The observed deviations between the three measurements largely fall within the uncertainty model.

\subsection{Sensor Optimization and Characterization}

We next explore the effect of changing the polarization of the probe and coupling beams (Fig.~\ref{Fig:Polarization}). In the experiments conducted here, our Rabi rates of the probe, coupling, and RF fields are $\Omega_p \simeq 2\pi \times 20$~MHz, $\Omega_c \simeq 2\pi \times 0.5$~MHz, and $\Omega_{RF} \simeq 2\pi \times 200$~MHz, respectively. Because $\Omega_{RF} \gg \Omega_c$ and $\Omega_{RF} \gg \Omega_p$, the RF field sets the quantization axis for the Rydberg atoms and is in the $\hat{\theta}$ direction. Thus, as we vary the angle of the RF field, the quantization axis changes and the RF electric field is oriented parallel to the PEC plate at $\theta = 0^{\circ}$ and  perpendicular to the plate at $\theta = 90^{\circ}$. When the probe and coupling beams are both linearly vertically polarized (Fig.~\ref{Fig:Polarization}a), their polarization axes align with the RF field only at $\theta = 0^{\circ}$. In the atomic reference frame, the probe and coupling polarizations thus change from $\pi$ polarized at $\theta = 0^{\circ}$ to $\sigma^{\pm}$ polarized at $\theta = 90^{\circ}$. 

The changing probe and coupling polarizations in the atomic reference frame as the RF angle is varied leads to two difficulties in determining the incoming angle. First, the magnitude of the central EIT peak that does not undergo AT splitting varies as a function of angle (Fig.~\ref{Fig:Polarization}b), which in turn reduces signal-to-noise ratio as $\theta \rightarrow 90^{\circ}$. We evaluate the fraction of the EIT signal remaining in the central peak by fitting three Lorentzians to each trace, and taking the area of the central peak fit divided by the area of all three fits. In the case when the probe and coupling are both linearly polarized, we see that $\sim50\%$ of the population does not undergo splitting. This effect is reduced when only one of the probe or coupling lasers are linearly polarized, but the best case is when both probe and coupling beams are circularly polarized. Second, the AT-split peaks further split into multiple sub-peaks which are not resolvable across all angles (Fig.~\ref{Fig:Polarization}b), leading to an increase in the uncertainty of $E_A/E_B$. Comparing the four combinations of probe and coupling polarizations, we see that linearly polarized probe and linearly polarized coupling deviate from other polarization cases in the response of $E_A/E_B$ to $\theta$ (Fig.~\ref{Fig:Polarization}e). Since circularly polarized probe and coupling maximize our signal and minimize our fitting uncertainty, we use this setup as the best combination.

The polarization effect between probe, coupling, and RF sources can be explained by examining the Cesium transition state diagrams in the hyperfine basis (Fig.~\ref{Fig:Polarization}f-g). The hyperfine transition diagrams were obtained from Alkali Rydberg Calculator (ARC) \cite{sibalic_arc_2017} and the allowed transitions are plotted as connected nodes with the dipole matrix element strength plotted by opacity \cite{Noah2024, PhysRevApplied.4.044015, schlossberger_zeeman-resolved_2024}. Consider first linearly polarized probe and coupling beams while the RF angle $\theta$ is varied. At $\theta = 0^{\circ}$, all sources are co-polarized in the $\hat{z}$ direction and only $\pi$ transitions occur (Fig.~\ref{Fig:Polarization}f). The dominant dipole elements through all transitions are in the low $m_F$ states and the population is driven towards the center with a low distribution in dipole moments. The subpeaks arising from this low distribution in dipole moments are not directly resolvable and convolve together into a wide peak (Fig.~\ref{Fig:Polarization}b). Now, at $\theta = 90^{\circ}$, the quantization axis set by the RF is polarized in the $\hat{x}$, so that the projection for the $\hat{z}$-polarized light into the $\hat{x}$ direction is given by a linear combination of $\hat{\sigma^{+}}$ and $\hat{\sigma^{-}}$ (Fig.~\ref{Fig:Polarization}g). This leads to two effects. First, we see strong transitions towards a trapped population at 52D\textsubscript{5/2} (F=6, $m_f = \pm 6$), which do not have allowed RF transitions. Therefore, this contributes to a central EIT peak that does not exhibit Autler-Townes splitting. Second, the transitions are driven towards a larger gradient in dipole moments at the 53P\textsubscript{3/2} state, which leads to multiple sub-peaks observed in the AT-splitting. In the case of circular probe and circular coupling optical fields, the projection of these light fields into the quantization axis is the same for all values of $\theta$, meaning that the hyperfine distribution will not change.  The light projects into a combination of $\pi$ and $\sigma^{\pm}$ transitions. From (Fig.~\ref{Fig:Polarization}a) we see that the majority of the Rydberg population undergoes AT-splitting, indicating that the $\pi$ transitions dominate throughout $\theta$, resulting in an easily resolvable AT-split peak with maximum signal-to-noise ratio.

In many applications including the phased array antenna configurations used in 5G, polar angle, $\theta$ (horizontal streerability),  is sufficient to locate the AoA of an incoming RF signal while the azimuthal angle, $\phi$ (vertical steerability), varies only slightly. Nonetheless, we test the extent to which $\theta$ can be determined for varying $\phi$ (Fig.~\ref{Fig:Phi}). The simulations indicate the monotonic behavior of $E_A/E_B$ versus $\theta$ still occurs for $\phi \leq 45^{\circ}$, although a slight variation in $E_A/E_B$ is seen as $\phi$ increases. The measured data observed a similar trend of monotonic determination of $\theta$ for $-22.5^{\circ} \leq \phi \leq 22.5^{\circ}$ (Fig. \ref{Fig:Phi}b). We note that in the case of $\phi = \pm 45^{\circ}$, reflections off the nearby optical mounts likely caused significant scattering of the incident fields leading to noise in $E_A/E_B$. Further experiments are needed to find if a pair of PEC-Rydberg cells oriented perpendicular to each other could locate the $(\theta, \phi)$ coordinates of an incoming RF-field.

To finish demonstrating the receiver characteristics in broad angular and polarization space, we next investigated the effect of changing RF polarization in uniquely determining the incoming AoA. In the data presented so far, the RF field was linearly polarized perpendicular to the major axis of the cylinder, $\chi = 0^{\circ}$. Changing the RF polarization changes the field patterns at varying $\theta$ along the laser beam paths as shown in Fig.~\ref{Fig:RFpolarization}a.  These conclusions from the simulations are observed in our experimental data. In particular, we see the sensitivity to $\theta$ is highest for $\chi = 0^{\circ}$, lower at $\chi = 45^{\circ}$, and insensitive to varying $\theta$ at $\chi = 90^{\circ}$. However, as a standalone measurement, $E_A/E_B$ becomes non-unique for an arbitrary unknown polarization (Fig.~\ref{Fig:RFpolarization}c). To break this degeneracy, we consider a secondary metric which determines the incoming polarization. 


Increasing $\chi$ from $0^{\circ}$ to $90^{\circ}$ leads to an increase in the atomic population not undergoing AT-splitting and occurs across all $\theta$ (Fig.~\ref{Fig:RFpolarization}b). This arises because when the RF polarization is parallel to the light propagation direction, the atoms only undergo $\sigma^{+}$ transitions and become trapped in the 52D\textsubscript{5/2} $(F=6, m_F = +6)$ state as described previously. Thus, by extracting two metrics from a measurement, the fraction in the central peak and the ratio $E_A/E_B$, we can measure the incoming AoA for RF polarizations between $\chi = 0^{\circ}$ and $45^{\circ}$.

\subsection{Future Improvements}

Having demonstrated the directional receiving characteristics of our PEC-Rydberg cell, we next investigated improvements that could be applied in transitioning the device towards field applications. In most laboratory experiments investigating Rydberg vapor cell physics with RF fields, the sensor is placed on a dielectric rod and surrounded by RF absorber -- thus acting as a free floating device. However, in real-world applications, the sensor would ideally be placed on or near metallic objects. Because our PEC-Rydberg sensor specifically exploits a conductor to measure standing wave fields, we hypothesized that we could still determine the incoming AoA when the sensor is mounted on a conducting surface. To test this, we wrapped a planar surface in multiple layers of foil tape and cut a hole for the Rydberg sensor. We manually positioned the metallic surface to lie nearly in the same xy-plane as the PEC plate (Fig.~\ref{Fig:Applications}). We observed a general trend of monotonically decreasing $E_A/E_B$ with increasing $\theta$, but note the relationship is not perfect. The cause of the deviations away from one-to-one mapping may be due to either secondary reflections off the optical mounts or the metallic plane not being perfectly co-planar with the PEC in the Rydberg vapor cell. Future optimization of the geometries could enable deployment for real-world measurements.  

In the data presented so far, we demonstrate the proof-of-principle of our Rydberg-PEC sensor in the frequency regime of field measurements, whereby a relatively strong RF field is applied in order to resolve the AT-split peaks from the zero field EIT linewidth. Alternatively in the amplitude regime, the EIT height near zero detuning can be measured for low applied RF field \cite{sedlacek_microwave_2012}. In general, this method requires power-stabilized optical fields. Although we did not use power-stabilization in our optical setup, we test the amplitude regime method to see if AoA measurements can still be determined. We apply field strengths of approximately 1.9~V/m at 4~GHz, 2~V/m at 5~GHz, and 1.2~V/m at 6~GHz and swept $\theta$ with $(\phi, \chi) = (0^{\circ}, 0^{\circ})$. In the case of 5.044~GHz and 5.711~GHz fields, we observe a trend of decreasing $E_A/E_B$ for increasing $\theta$. However, in the case of 4.228~GHz the ratio $E_A/E_B$ increased from $\theta=0^{\circ}$ to $\theta=20^{\circ}$, preventing the sensor from operating successfully across all $\theta$. Future efforts employing power stabilization are expected to improve the minimum detectable fields.

\section{Discussion \label{sec:Discussion}}

We demonstrated a method of determining the angle-of-arrival (AoA) of an incident radio-frequency (RF) field based on amplitude only measurements with a Rydberg atom sensor. Our method addresses limitations in AoA directional finding of both traditional and photonic-based antenna systems, which we discuss below.


Traditional antenna methods for AoA directional finding include  phase, amplitude and time delay. Current methods for amplitude only AoA finding include mechanically actuated systems \cite{hood2010estimating}, multiple number of squinted high gain antennas \cite{al-tarifi_amplitude-only_2016}, or wavelength-sized monopole arrays utilizing machine learning algorithms \cite{friedrichs2023angle}. These methods feature polar angular resolution of $\sim$1 to 8$^{\circ}$, comparable to the 4.1$^{\circ}$ reported here. Although these conventional antenna systems can meet specific demands, 
no system can simultaneously maximize sampling rate, angular resolution, and bandwidth. In contrast, our Rydberg sensor has ultrawideband tunability \cite{meyer2020assessment} and operates at subwavelength scales. Furthermore, the radar cross section of our PEC-Rydberg sensor is much less than the conventional antenna systems and hence could play an important role in tactical/military applications.

Previous AoA measurements with atomic Rydberg sensors required determining the relative phase of two measurements of the incident field  \cite{AOA, 10304615, yan2023three}. This approach requires the use of an additional RF field, i.e. a local oscillator (LO), to provide a phase reference. Our PEC-Rydberg cell removes the LO requirement for AoA detection and is independent of the incoming field strength within the acceptable limits. The angular resolution of our approach, 4.1$^{\circ}$, is similar to the LO-based approaches, $\sim2^{\circ}$ \cite{AOA, 10304615}. Interestingly, Yan et al. (2023) used 3 beams to locate the polar and azimuthal angles of the signal source in three-dimensions \cite{yan2023three}. We believe that by adopting a second PEC-Rydberg cell orthogonal to the first, our method would also enable $(\theta, \phi)$ localization.


RF scattering from vapor cell materials, like glass, can lead to non-uniform RF field distributions  \cite{rotunno_investigating_2023}, which can significantly impact AoA estimation if not mitigated properly \cite{richardson2024study}.  However, the structure used in this work specifically exploits a PEC plate to form standing patterns within the cell. The deformations of the field due to the geometry and materials of the vapor cell are minimal compared to the standing wave field structure and can be captured in the calibration of a given cell.

Since our method does not require absolute field strength measurements, it is only dependent on the ratio of two fields, many of the uncertainties associated with making SI-traceable field strength methods do not apply here \cite{simons2018uncertainties}. However, because our method utilizes a calibration, other Type~B uncertainties can arise. The dominant Type~B uncertainty in our measurement results is due to the beam distances relative to the conducting plate. We assume Beam A and B can be located to within $\pm$100~$\mu$m by finding the center of the beam position shown in Fig.~\ref{Fig:ExperimentalSetup}c. In this case, the largest deviation to the ratio occurs for Beam B moving away from the plate by 100~$\mu$m and Beam A moving towards the plate by 100~$\mu$m. We take the Type~B beam deviation as a rectangular distribution so that the uncertainty is the max deviation over $\sqrt{3}$. In this case, we have an AoA uncertainty of $\sim4^{\circ}$ at $\theta = 0^{\circ}$. This uncertainty decreases as the AoA, $\theta$, increases because the slope of the field with respect to the distance from the plate decreases as can be seen in Fig.~\ref{Fig:Simulation}g, with $\sim1^{\circ}$ uncertainty at $\theta = 30^{\circ}$.

    

\section{Methods}\label{Methods}

\subsection{Numerical Simulation\label{sec:Simulation}}

We model the E-field response for plane waves incident on a glass vapor cell using a commercial finite element solver. The geometry is shown in Fig.\ref{Fig:Intro}. A $25$ mm x $75.5$ mm glass vapor cell of thickness $0.95$ mm and dielectric constant 5.5 lies with it's axis along the $\hat{y} $ direction. A perfect electrically conducting plane (PEC) lies in the cell in the xy-plane with an offset of  1 mm in $-\hat{z}$ direction to replicate the fabricated cell.  The source of the RF field is a linearly polarized plane wave with frequency from 500~MHz to 10~GHz and the incoming direction, $\hat{k}$ is varied as a function of $\theta$ and $\phi$. The lower frequency limit was chosen because Rydberg vapor cells begin experiencing charge density screening in the 100s of MHz, and the upper limit as the AoA monotonic behavior was seen to begin breaking down. The polarization of the plane wave is parameterized to study its effect on the field distributions. Perfectly matched layer (PML) boundary conditions at appropriate distances from the cell are used in the simulation. Two lines parallel to the PEC plane indicate field sampling locations at distances $z_A$ and $z_B$. 

\subsection{Optical Setup \label{sec:Optical Setup}}

We designed an optical setup for a Rydberg atomic E-field sensor to experimentally test the simulation results that there is a one-to-one mapping of the ratio of field strengths versus incoming AoA. We measure field strengths through electromagnetically induced transparency (EIT) and resonant Autler-Townes splitting in the Cesium, \textsuperscript{133}Cs, transition (Fig.\ref{Fig:Transitions}) \cite{holloway_broadband_2014}. Briefly, EIT is a process whereby the probe and coupling lasers create a quantum superposition of the excited states rendering the medium transparent to both lasers. The Autler-Townes effect then occurs when RF radiation incident on the atom, which is resonant to a transition between two states of the atom, perturbs and splits each of these energy levels. The split of these energy states can in turn be used to measure the applied RF field. 

This splitting of the EIT spectrum is directly proportional to the applied RF E-field amplitude \cite{fleischhauer_electromagnetically_2005}. By measuring the optical frequency difference of this splitting  we get a direct measurement of the RF E-field strength from the following expression, \cite{holloway_atom-based_2017, holloway_broadband_2014, sedlacek_microwave_2012}
\begin{equation}
    |E| = 2\pi \cfrac{\hbar}{\varrho} \Delta f_o,
\end{equation}
where $\hbar$ is the Planck constant, $\varrho$ is the atomic dipole moment of the RF transition, and $\Delta f_o$ is the Autler-Townes frequency splitting.

We designed a portable optical setup capable of controlling the polarization and separation of two probe beams and two coupling beams in the PEC test cell (Fig.~\ref{Fig:ExperimentSetup_a}). The probe beams had a power of  approximately 1~mW $\pm$~0.4~mW and $1/e^2$ beam diameter of 2.1~mm at the cell. The coupling beams had a power of approximately 5~mW $\pm$~0.3~mW and beam diameter of 1.2~mm. As the method described in this paper takes only the ratio of the field strengths at two locations, there is virtually no requirement for power and frequency stabilization. The separation between the beams was 7~mm as shown in  Fig.~\ref{Fig:ExperimentSetup_a}. The reference cell gives a frequency reference between the main 6SP\textsubscript{3/2} (F=5) $\rightarrow$ nD\textsubscript{5/2} and the sub-peak 6SP\textsubscript{3/2} (F=4) $\rightarrow$ nD\textsubscript{5/2} \cite{steck_cesium_nodate, sibalic_arc_2017}. The probe and coupling beam's polarizations were controlled by a $\lambda/2$ wave-plate to align all 4 beams into vertical polarization followed by a pair of $\lambda/4$ wave plates to change from linear to circular polarizations. RF absorber separates all the optical elements from the test cell except for three mirrors with mounts. The PEC test cell was placed on a dielectric rod. To verify the absorber did not affect the field measurements, we added secondary pieces of absorber and observed no difference in the Autler-Townes splitting for varying $\theta$. 

We measured field responses through the WR187 waveguide band of 3.95 to 5.85~GHz. Specifically, we investigate the transitions: $n$ = 50, $f_C=$588.36046~THz, $f_{RF}=5.711$~GHz; $n$ = 52, $f_C=$588.47567~THz, $f_{RF}$ = 5.044~GHz; and $n$ = 57, $f_C=$588.71027~THz, $f_{RF}$ = 4.228~GHz. The Rayleigh distance, $Z$, for the WR187 open ended waveguide antenna at 5~GHz is given by $Z = 2D^2/\lambda_{RF}$ = 9.2~cm, where $D$ is the diagonal distance of the WR187 open ended waveguide antenna. The 3 dB E-plane beamwidth is $\sim 78^{\circ}$ \cite{noauthor_187ewg7_nodate}. As a balance between obtaining far-field measurements versus generating reflections off the nearby optical mounts due to the relatively wide beamwidth, we performed all measurements at a distance of 20 cm between the waveguide open end and the Rydberg vapor cell.

\subsection{LAPS Facility \label{sec:Robo_Facility}}


We conducted measurements using the Large Antenna Positioning System (LAPS), in which a 7 Degree of Freedom (DoF) 
subset of the positioning system composed of a 6 DoF robotic arm mounted to a linear rail was used to permit flexible positioning in the chamber. The position and orientation of the waveguide attached to the robot end effector was calibrated using laser tracker measurements \cite{Moser2024}. The motion coordinate system used for positioning the robot within the chamber was established by replacing the vapor cell with a Spherically Mounted Retroreflector (SMR) \cite{Muralikrishnan2016}. A laser tracker measurement of this location is used as the origin, and principle axes were established through measurement of the surface and leading edge of the optical breadboard with precision SMR nests. Angle-of-arrival measurement plans (the set of planewave incident angles relative to the vapor cell coordinate system) were created by first generating the desired transforms between the vapor cell coordinate system and the waveguide corresponding to the desired $\theta$, $\phi$, and $\chi$ coordinates. Then, robot inverse kinematics were solved for each transform to produce unique joint configurations that were sequentially streamed to the robot controller during measurement.

\section*{Acknowledgments}
This research was developed with funding from the Defense Advanced Research Projects Agency (DARPA) and was supported by NIST under the NIST-on-a-Chip program. The views, opinions and/or findings expressed are those of the author and should not be interpreted as representing the official views or policies of the Department of Defense or the U.S. Government. Distribution Statement "A" (Approved for Public Release, Distribution Unlimited).  A contribution of the U.S. government, this work is not subject to copyright in the United States. The computational results in this work were made possible by the Baker-Jarvis high performance computer cluster in the Communications Technology Laboratory at NIST and supported by NIST's Research Services Office. Certain commercial equipment, instruments, or materials are identified in this paper in order to specify the experimental procedure adequately. Such identification is not intended to imply recommendation or endorsement by the National Institute of Standards and Technology, nor is it intended to imply that the materials or equipment identified are necessarily the best available for the purpose.

\subsection*{Conflict of Interest}
\vspace{-3mm}
The authors have no conflicts to disclose.

\vspace{-3mm}
\subsection*{Data Availability Statement}
\vspace{-3mm}
All data presented in this paper is available at https://doi.org/10.18434/mds2-3609.

\newpage

\bibliography{main}
\end{document}